\documentclass[preprint,superscriptaddress,aps]{revtex4}

\usepackage{graphicx}
\usepackage{dcolumn}
\usepackage{bm}

\begin{document}

\title{New quantum state formed by highly concentrated protons in superconducting palladium hydride} 

\author{Ryoma Kato}
\affiliation{Department of Applied Quantum Physics, Kyushu University, Motooka, Fukuoka 819-0395, Japan}
\author{Ten-ichiro Yoshida}
\affiliation{Department of Applied Quantum Physics, Kyushu University, Motooka, Fukuoka 819-0395, Japan}
\author{Riku Iimori}
\affiliation{Department of Physics, Kyushu University, Motooka, Fukuoka 819-0395, Japan}
\author{Tai Zizhou}
\affiliation{Department of Applied Quantum Physics, Kyushu University, Motooka, Fukuoka 819-0395, Japan}
\author{Masanobu Shiga}
\affiliation{Department of Applied Quantum Physics, Kyushu University, Motooka, Fukuoka 819-0395, Japan}
\author{Yuji Inagaki}
\affiliation{Institute for the Advancement of Higher Education, Okayama University of Science,
Ridaicho, Kita-ku, Okayama 700-0005, Japan}
\author{Takashi Kimura}
\affiliation{Department of Physics, Kyushu University, Motooka, Fukuoka 819-0395, Japan}
\author{Koichiro Ienaga}
\affiliation{Graduate School of Sciences and Technology for Innovation, Yamaguchi University, Ube, Yamaguchi, 755-8611, Japan}
\author{Tatsuya Kawae}
\email{t.kawae.122@m.kyushu-u.ac.jp}
\affiliation{Department of Applied Quantum Physics, Kyushu University, Motooka, Fukuoka 819-0395, Japan}


\begin{abstract}
Hydrogen exhibits quantum phenomena, such as tunneling in materials. According to theory, the quantum properties of hydrogen change significantly in superconductors due to the emergence of an energy gap on the Fermi surface, which reduces the interaction between hydrogen nucleus (i.e., proton) and conduction electrons. This reduction is predicted to enhance the tunneling probability of protons. Here, we report the double transitions of the electrical resistivity in high-quality palladium hydride (PdH$_{x}$) and deuteride (PdD$_{x}$) prepared by low-temperature absorption below $T$ $=$ 180 K. After a sharp drop in the resistivity at $T$ $\sim$ 2 K owing to the superconducting transition of PdH(D)$_{x}$, a large residual resistivity remained. Additionally, the resistivity dropped to zero below $T$ = 1 K. The experimental results suggest that the quantum tunneling of highly concentrated protons (deuterons) in the superconducting state is responsible for the observed features: the residual resistivity arises from the weakening of the global coherence of superconductivity owing to the tunneling motion of protons (deuterons), while the zero resistivity is caused by long-range ordering of the protons (deuterons). This system offers a new platform for investigating the quantum many-body properties of tunneling particles.
\end{abstract}


\maketitle{}

\section{INTRODUCTION}
Hydrogen (H), with the smallest atomic mass, exhibits various quantum phenomena in materials at low temperatures. In transition metals, where H atoms occupy interstitial sites, they migrate by quantum tunneling~\cite{1,2,3,4} in a sea of conduction electrons at low temperatures, as shown in Fig. 1(a)~\cite{5,6,7}. Theories indicate that this situation varies significantly in the superconducting state~\cite{8,9,10,11,12,13}. An energy gap appears on the Fermi surface, which suppresses the interaction between hydrogen nucleus (i.e., proton) and conduction electrons, as illustrated in Figs. 1(b) and 1(c), and then influences the quantum behavior of the H atoms. For example, the tunneling probability of a proton is predicted to increase exponentially with decreasing temperature. However, experimentally, the variation in the tunneling features of H at the interstitial site due to the superconducting transition has not been revealed. Notably, the tunneling diffusion of muons with a mass of approximately 1/9 that of a proton has been confirmed to increase markedly after the superconducting transition~\cite{14}, which is in good agreement with the theoretical prediction. 

Palladium-hydride (PdH$_{x}$) is suitable not only for studying the role of H in the occurrence of superconductivity in hydride systems but also for exploring the quantum behaviors of highly concentrated H in superconductors. This is because the superconducting transition of PdHx occurs for the H concentration $x$ $\gtrsim$ 0.72 under ambient pressure~\cite{15,16}. Moreover, the transition temperature Tc increases with H concentration17) and reaches $T_{c}$ of ~9 K for a stoichiometric concentration of $x$ = 1~\cite{18,19}. This facilitates various experimental investigations, in contrast to room-temperature superconductors in H-rich materials discovered at extremely high pressures~\cite{20,21,22}. However, the superconducting properties of PdH$_{x}$ have not been sufficiently investigated so far owing to the low quality of PdH$_{x}$ samples, which results in a broadening of the superconducting transition. 

We successfully prepared a high-quality sample of PdH$_{x}$ using a low-temperature H absorption method, in which H atoms were loaded into Pd under an H$_{2}$ atmosphere below $T$ = 200 K~\cite{19,23}. Subsequently, the PdH$_{x}$ sample was cooled to a low temperature for the measurements without H desorption. For PdH$_{x}$ powders with diameters of 1–2 ${\mu}$m, the superconducting volume fraction reached $\sim$1, indicating that samples with a uniform H concentration can be obtained~\cite{19,23} Using this method, we studied the temperature dependence of the resistivity of the PdH$_{x}$ films~\cite{24,25}. By loading H atoms in a Pd film with a thickness of 100 nm for approximately 20 days, a PdH$_{x}$ film with a superconducting transition width below 0.1 K (where $T_{c}$ $\sim$ 1.1 K) was achieved, which was considerably narrower than that of the previous measurements. However, a residual resistance of unknown origin remained. 

In this paper, we report the discovery of double transitions in the resistivities of PdH$_{x}$ and PdD$_{x}$ films. As in our previous studies, the residual resistivity appears below the superconducting transition. In addition, zero resistivity is observed below the second transition in the lowest-temperature region. The observed features can be interpreted as tunneling motion and the ordering of highly concentrated protons or deuterons in the superconducting state.

\section{EXPERIMENTAL}
We studied the electrical resistivity of two types of Pd films with a thickness of 100 nm, which is considerably larger than the coherence length of superconducting PdH$_{x}$ ~\cite{16,23}. One is a Pd film deposited on an Au electrode, which enables the absorption of H atoms in the Pd film without obstructing the electrodes The other is that Cu electrodes are deposited on the Pd film, which was prepared for comparison with the Au electrode results. The detail of the sample synthesis is presented in Appendix.

The resistivity measurements were conducted in the same manner as in the previous measurements~\cite{25}. The Pd film was mounted on a Cu plate installed in a homemade $^{3}$He insert that was attached to a Quantum Design MPMS SQUID magnetometer~\cite{26}. The resistance of the film was measured by a four-wire method using an LR700 AC resistance bridge down to a temperature of $\sim$0.6 K, from which the resistivity was estimated. The RuO$_{2}$  thermometer was attached to the Cu plate~\cite{25}. A magnetic field was applied parallel to the PdH$_{x}$  film along the current direction. It should be noted that the field deviates from the parallel direction because a thin Cu plate is used as the sample holder, which deforms easily during the mounting of the holder onto the $^{3}$He insert. The experimental procedure is detailed in Appendix.

\section{RESULTS}
\subsection {Temperature dependence of resistivity in PdH$_{x}$  films}
In the previous measurements, the electrodes were attached to a Pd film using silver (Ag) paste~\cite{24,25}.
However, in this setup, H absorption may be prevented by the electrode, resulting in H deficiency. Consequently, residual resistivity could appear below the superconducting transition temperature.  Therefore, in the present study, we prepared new Pd films with a thickness of 100 nm and width of 100 ${\mu}$m on Au electrodes (Fig. A-1(a)), which enabled the absorption of H atoms in the Pd film without obstruction by the electrodes. Figure 2(a) shows the temperature dependence of resistivity below 5 K for the PdH$_{x}$  film, where H absorption was performed at $T$ = 150 K. 

The resistivity is constant above the superconducting transition temperature, dropping sharply below $T_{c1}$  = 2.14 K; the transition finishes at 2.0 K, reflecting a uniform H distribution in the film. The H concentration is estimated to be $x$ = 0.865 from the transition temperature $T_{c1}$ = 2.14 K~\cite{17}. Despite the sharp drop, a large residual resistivity remains, exhibiting a gradual decrease with decreasing the temperature, as shown in the inset of Fig. 2(a). These features are qualitatively the same as those in the previous measurements with electrodes on a Pd film~\cite{24,25}. Notably, the second transition with a steep drop in resistivity occurs at $T_{c2}$ = 0.78 K, resulting in zero resistivity. It is also significant that the ratio of residual resistivity below $T_{c1}$ to the resistivity in the normal state is much larger than that of the previous measurements with electrodes on the Pd film. This indicates that the residual resistivity is not caused by H deficiency due to the electrodes covering the Pd film.

To clarify the origin of the residual resistivity and second transition, we performed two measurements, varying (i) the H concentration and (ii) the size of the film (details of measurement (ii) are described in Appendix “Resistivity measurements in 10 ${\mu}$m wide film”). To increase the H concentration of the film, the temperature of the PdH0.865 film was raised to $T$ = 150 K, and the film was exposed to an H$_{2}$ gas atmosphere (the procedure is outlined in Appendix “Procedure of hydrogen absorption into Pd film”). After reabsorption, the temperature dependence of the resistivity was measured again, as depicted by the red plots in Fig. 2(b). The normal resistivity before the superconducting transition decreases owing to the increase in H concentration~\cite{27}. Moreover, the two transition temperatures are increased to $T_{c1}$ = 2.31 K, corresponding to $x$ = 0.870, and $T_{c2}$ = 0.86 K. Subsequently, the H concentration was decreased by increasing the temperature to $T$ = 180 K to desorb H from the film, as shown by the green plots in Fig. 2(b). The superconducting transition temperature decreases to $T_{c1}$ = 1.28 K, corresponding to $x$ = 0.834, whereas the second transition is invisible. The H concentration of the film was increased again by reabsorbing H into the film, raising the two transition temperatures to $T_{c1}$ = 1.85 K at $x$ = 0.855 and $T_{c2}$ = 0.69 K, as depicted by black plots in Fig. 2(b).

The concentration dependences of $T_{c1}$ and $T_{c2}$ are plotted in Fig. 2(d). Through H absorption and desorption, $T_{c2}$ appears to change almost linearly as a function of the H concentration. $T_{c2}$ at $x$ = 0.834 is not visible because it is estimated to be lower than the minimum temperature in the experiment. Evidently, the slope of $T_{c2}$ is different from that of $T_{c1}$. Additionally, the magnitude of the residual resistivity above $T_{c2}$ is approximately the same for all concentrations, as shown in Fig. 2(c), in contrast to the marked variation in the resistivity in the normal region, which depends remarkably on the H concentration~\cite{27}. These results strongly suggest that the origin of $T_{c2}$ is not caused by H deficiency.

Similar characteristics are observed in the PdH$_{x}$ film with a width of 10 ${\mu}$m, which is one-tenth the size of the current film (Fig. A-3). This indicates that the residual resistivity and the second transition do not depend on the sample size.

We examined the influence of Au electrodes covered with a Pd film. Since Au can form an Au-Pd alloy~\cite{28,29}, it might affect the superconducting transition of the Pd film. To assess this possibility, we measured the resistivity of a PdH$_{x}$ film with a different electrode structure, in which Cu electrodes were deposited on the Pd film illustrated in Fig. A-1(b), meaning that the electrodes were not covered by the Pd film. Details of this measurement are given in Appendix “Resistivity in Pd film with Cu electrodes”. The qualitative features are consistent with those obtained for the Au electrodes. These results demonstrate that the emergence of the residual resistivity and the second transition do not originate from the influence of the Au electrodes but from the intrinsic features of superconducting PdH$_{x}$.

\subsection{Magnetic field dependence of superconducting transition in PdH$_{x}$ film}
Figures 3(a) and 3(b) show the magnetic field dependence of the resistivity around $T_{c1}$ and $T_{c2}$, respectively, in the PdH$_{0.870}$ film. As the magnetic field increases, the superconducting transition at $T_{c1}$ is suppressed with broadening the transition curve, as in the case of conventional superconducting transitions. In contrast, although $T_{c2}$ is also suppressed, the sharp drop in resistivity is maintained up to $H$ = 800 Oe, which is the maximum field at which $T_{c2}$ can be observed. For instance, at $H$ = 0 Oe, the resistivity starts to deviate at 0.86 K and decreases to zero at 0.83 K, while at $H$ = 800 Oe, these values are at 0.59 K and 0.56 K, respectively. These characteristics are clearly illustrated in the phase diagram of $T_{c1}$ and $T_{c2}$ in Fig. 3(d), where no broadening of $T_{c2}$ is observed up to $H$ = 800 Oe. These results suggest that the transition at $T_{c2}$ does not originate from the superconducting transition of PdH$_{x}$. 

Figure 3(c) plots the field dependence of the resistivity as the magnetic field increases at $T$ = 0.6 K in PdH$_{0.870}$ film. As the magnetic field increases, the resistivity rises steeply at $H$ = 770 Oe and reaches the value of the residual resistivity at $H$ = 800 Oe, as shown in the inset. Subsequently, the resistivity is constant up to $\sim$2500 Oe, which is followed by a gradual increase and rises sharply at approximately $H$ = 2700 Oe. This value is in reasonable agreement with the upper critical field observed in the magnetization curve of PdH$_{0.870}$, where the measurements were performed in the powder sample with a diameter of 1$\sim$2  ${\mu}$m~\cite{16} However, a transition at $T_{c2}$ cannot be observed in the magnetization curve. This also supports that the transition at $T_{c2}$ does not originate from the superconducting transition of PdH$_{x}$.

\subsection{Superconducting transition in PdD$_{x}$ film }
Finally, we present the resistivity of a palladium deuteride (PdD$_{x}$) film, where the Pd film has a thickness of 100 nm and a width of 100 ${\mu}$m, deposited on the Au electrodes (Fig. A-1(a)). The temperature and magnetic dependences of the resistivity is shown in Figs. 4(a) and 4(b), respectively. The overall features are similar to those observed in the PdH$_{x}$ measurements. The resistivity decreases sharply at $T_{c1}$ = 1.47 K owing to the superconducting transition, from which the deuterium concentration is estimated to be $x$ = 0.795~\cite{17}. A large residual resistivity remains below $T_{c1}$, and a steep drop in resistivity, attributed to the second transition, occurs at $T_{c2}$ $\sim$ 0.6 K. The magnitude of the residual resistivity above $T_{c2}$ is approximately the same as that in the PdH$_{x}$ film with the same sample size shown in Figs. 2(a)-2(c), despite the fact that the mass of deuterium is twice that of a proton. As the magnetic field increases, the superconducting transition at  $T_{c1}$ is suppressed with broadening the transition curve.

However, a close examination of the temperature dependence demonstrates a clear difference between the PdH$_{x}$ and PdD$_{x}$ results. In the PdD$_{0.795}$ film, a rectangular-shaped anomaly appears between $T$ = 0.95 and 1.2 K, as presented in the inset of Fig. 4(a). As the magnetic field increases, this anomaly shifts to lower temperatures (Fig. 4(c)), where  $T_{c1}$ decreases with the broadening of the resistivity curve, as shown in Fig. 4(b). Since a rectangular anomaly is not detected in the PdH$_{x}$ film, it is reasonable to consider that an isotope effect contributes to the properties of the residual-resistivity state in superconducting PdD$_{x}$~\cite{30}.

\section{DISCUSSION}
As described above, the residual resistivity in PdH(D)$_{x}$ films cannot be explained by H deficiency or the influence of the electrodes, indicating that the emergence of the residual resistivity followed by the second transition is an intrinsic feature of a high-quality PdH(D)$_{x}$ film. These results strongly suggest that protons (deuterons) immersed in the superconducting state play a pivotal role in the observed features. Similar behaviors, that is, residual resistivity followed by a zero-resistivity transition, have been observed in several systems, for example, vortex liquid-lattice transition in a superconducto~\cite{31,32}, and charge density wave (CDW) coexisting with superconductivity~\cite{33}, in which the resistive state exists in the superconducting state. In the vortex system, the motion of the vortices disrupts the global coherence of superconductivity, generating resistivity in the vortex liquid state, whereas freezing the motion leads to zero resistivity. Notably, a sharp anomaly in resistivity emerges at the vortex liquid-lattice transition~\cite{32}. 

Similarly, the liquid-like tunneling motion of protons is expected to disrupt the global coherence of superconductivity. According to theories~\cite{9,34,35}, the energy dissipation of tunneling protons is inevitable due to the interaction with the electron and lattice systems, causing phase fluctuation of the superconducting wave function. Consequently, the global coherence of the superconducting state is disrupted, leading to residual resistivity. Conversely, when protons or deuterons order, the effective mass of the tunneling particle increases, dramatically decreasing the tunneling probability. This results in the observation of zero resistivity. This scenario is consistent with the fact that the anomaly at  $T_{c2}$ cannot be observed in the magnetization measurements, which are sensitive to vortex penetration. 

The residual resistivity of the PdH$_{x}$ films with Au electrodes is considerably larger than that of the Cu and Ag electrodes presented in our previous paper~\cite{25}. This difference could be attributed to variations in tunneling probability, which are influenced by the energy difference between potential wells at the tunneling site, as illustrated in Fig. 1(a)~\cite{9}. Attaching electrodes after film deposition can degrade the homogeneity of the Pd film. This degradation increases the potential difference (${\Delta}{\varepsilon}$) between the two wells for H tunneling, thereby suppressing tunneling probability by widening the potential difference at the tunneling sites~\cite{9}. Accordingly, the residual resistivity is lower for Cu and Ag electrodes than Au electrodes.

Finally, we consider the nature of the ordering below $T_{c2}$. PdH$_{x}$ system is known to undergo a glass-like transition at approximately 50 K~\cite{36}. Below $T_{c1}$, the protons move in a liquid-like manner owing to an increase in the tunneling probability, which should lead to ordering at lower temperatures. Therefore, one possibility is a structural ordering that stabilizes the proton or deuteron positions. Indeed, the lack of a clear correlation between $T_{c1}$ and $T_{c2}$ across different PdH$_{x}$ films, as shown in Fig. 2(d), supports this idea. However, simple structural ordering cannot explain the rectangular anomaly in the resistivity of PdD$_{x}$. 

Another possibility is the magnetic ordering or condensation of protons and deuterons. The emergence of the rectangular anomaly, detected exclusively in the PdD$_{x}$ film, indicates that the statistical differences arising from the nuclear spin degree of freedom ($I$ = 1/2 for protons, $I$ = 1 for deuterons) influence the tunneling behavior of PdH$_{x}$ and PdD$_{x}$. This effect becomes particularly significant in the highly concentrated H or D regime, where superconductivity occurs for $x$ ${\gtrsim}$ 0.75. Since nuclear spin contributions typically manifest at temperatures well below $T$ = 1 K~\cite{37},  the emergence of the rectangular anomaly around $T$ = 1 K suggests that nuclear contributions are enhanced by quantum many-body interactions between protons (or deuterons) and the electron and/or lattice systems. These interactions may influence the ordering process as well as the tunneling behavior. Clearly, further experimental and theoretical studies are needed to gain a deeper understanding of the tunneling dynamics and ordering mechanisms.

\section{CONCLUSION}
We observed double transitions in the electrical resistivities of the PdH(D)$_{x}$ films. Despite a sharp superconducting transition at $T_{c1}$, a large residual resistivity persists below $T_{c1}$, which then drops to zero at $T_{c2}$. The experimental results are explained by assuming that the tunneling probability of protons (deuterons) is enhanced by the formation of a superconducting gap, as suggested by theories~\cite{8,9,10,11,12,13} The tunneling motion of protons (deuterons) disrupts the global coherence of superconductivity, causing residual resistivity. An ordering of the tunneling motion at $T_{c2}$ increases the effective mass, suppressing the tunneling probability and leading to zero resistivity. These findings suggest that tunneling protons (deuterons) in superconductors form novel quantum states, opening a pathway for studying the quantum many-body properties of tunneling particles.

\section{ACKNOWLEDGMENTS}
We specially thank Profs. Matsumoto, Shirahama and Maruyama for fruitful discussion. We also thank Professor Yamamuro for providing information about the low-temperature H absorption into Pd. We also thank Mrs. Hasuo and Yamaguchi for their technical help. This work was supported by JSPS KAKENHI Grant Numbers JP23K17763 and JP21H01605, Nippon Sheet Glass Foundation for Materials Science and Engineering, and Murata Science Foundation.

\section{Appendix}
\subsection{Synthesis of Pd film}

 We studied the electrical resistivity of two types of Pd films with a thickness of 100 nm, which is considerably larger than the coherence length of superconducting PdH$_{x}$~\cite{16,23}. One is a Pd film deposited on an Au electrode, which enables the absorption of H atoms in the Pd film without obstructing the electrodes, as shown in Fig. A-1(a). Au electrodes with thickness of 20  nm and width of 50  ${\mu}$m were prepared on a SiO$_{2}$/Si substrate by electron beam evaporation. To adhere the Au electrodes to the substrate firmly, a 5-nm Ti layer was deposited below the Au electrodes. Subsequently, a Pd film with a thickness of 100 nm and width of 100  ${\mu}$m was deposited on the substrate with Au electrodes by magnetron sputtering under a base pressure of 5.5×10$^{-6}$ Pa. The distance between the two voltage electrodes was 300  ${\mu}$m. In addition, a Pd film with a thickness of 100 nm and a width of 10  ${\mu}$m, deposited on Au electrodes, was prepared to reveal the influence of the film size.

The other film with Cu electrodes is shown in Fig. A-1(b). This sample was prepared for comparison with the Au electrode results. First, a Pd film with a thickness of 100 nm and a width of 100  ${\mu}$m was deposited on a SiO$_{2}$/Si substrate by magnetron sputtering under a base pressure of 5.5×10$^{-6}$ Pa. Subsequently, Cu electrodes with a thickness of 100 nm and width of 50  ${\mu}$m were deposited on the Pd film by thermal evaporation. The distance between the two electrodes was 300  ${\mu}$m.

\subsection{Low-temperature hydrogen absorption method}
H absorption was performed at temperatures below $T$ = 180 K. As shown in Fig. A-2, the H concentration $x$ in Pd increases as the temperature decreases. For example, when metallic Pd is exposed to an H$_{2}$ pressure of 0.1 MPa at $T$ = 200 K, the H concentration in Pd reaches approximately $x$ $\sim$0.8, which exceeds the critical concentration required for superconductivity in PdH$_{x}$. Using this low-temperature H absorption method, we conducted H charging in Pd metal and observed the superconducting transition through in-situ magnetization or resistivity measurements down to $T$ $\sim$0.6 K. In this study, we prepared PdH(D)$_{x}$ films by varying the absorption temperature; the H$_{2}$ (D$_{2}$) pressure was $\sim$0.25 MPa.

\subsection{Procedure of hydrogen absorption into Pd film}
The H absorption into the Pd film was conducted using the following procedure: Before cooling the sample for H absorption, several activation processes were performed at $T$ = 320 K by exposing the film to H$_{2}$ gas and evacuating it to remove surface contaminants and rearrange the surface structure to increase the H-absorption speed. After activation, the Pd film was cooled to $T$ = 180 K, and H$_{2}$ gas at the pressure of $\sim$0.25 MPa was introduced into the experimental chamber. With increasing H content in the film, the resistance varies remarkably~\cite{25,27,38,39}. 

The resistance variation approaches constant, demonstrating that the H content in the film is close to the equilibrium concentration at $\sim$0.25 MPa and 180 K. Thereafter, the temperature was decreased to 170 K for further H absorption. We repeated the above procedure in increments of 10 K. After H absorption at $T$ = 150 K, the temperature of the PdH$_{x}$ film decreased to 20 K at a cooling rate of 1 K/min to minimize the risk of sample breakage and H desorption. After cooling, H$_{2}$ gas was evacuated, and $^{3}$He gas was introduced into the experimental chamber, allowing further cooling for the measurements. The temperature dependence of the resistivity of the PdH$_{x}$ film obtained using this procedure is shown in Fig. 2(a). The H concentration is estimated to be $x$ = 0.865, based on the superconducting transition temperature.

The H reabsorption was performed as follows. After the resistivity measurements in PdH$_{0.865}$ film shown in Fig. 2(a), the temperature was increased to 150 K. Then, H$_{2}$ gas at $\sim$0.27MPa, higher than that used for the absorption at $x$ = 0.865, was introduced in the chamber to reabsorb H atoms into the film. After reabsorption was completed, the temperature was lowered again for the resistivity measurements. The results are plotted at $x$ = 0.870, which is estimated from the transition temperature $T_{c1}$ = 2.31 K. This procedure was repeated to change the H concentration. For H desorption, the temperature was raised to 180 K, and the resistivity was measured. The results are plotted at $x$ = 0.834, where the second transition is not visible. Afterward, the H concentration increased again upon loading H atoms at 150 K. The results are plotted at $x$ = 0.855.

\subsection{Resistivity in 10-${\mu}$m wide film}
To clarify whether the second transition at $T_{c2}$ originates from H deficiency, we performed resistivity measurements using a Pd film with a width of 10 ${\mu}$m, which is one-tenth smaller than that used in the measurements in Figs. 2(a) and 2(b). The thickness and distance between the two voltage electrodes are 100 nm and 300 ${\mu}$m, respectively. The arrangement of the Au electrodes is the same as that in Fig. A-1(a).

 The temperature dependence of the resistivity is plotted in Fig. A-3, where the H absorption is performed at $T$ = 150 K. The resistivity decreases sharply below $T_{c1}$ = 1.70 K, and the transition finishes at 1.62 K. The H concentration is $x$ = 0.850 at $T_{c1}$ = 1.70 K. As in the case of the PdH$_{x}$ film with a width of 100 ${\mu}$m, a large residual resistivity remains, which decreases gradually with decreasing the temperature, as shown in the inset. Moreover, the resistivity drops steeply at $T_{c2}$ = 0.83 K, leading to zero resistivity. These features agree well with the results in the Pd film with the width of 100 ${\mu}$m shown in Figs. 2(a) and 2(b). Meanwhile, $T_{c2}$ appears at a higher temperature compared to the results for the 100-${\mu}$m width, although $T_{c1}$ is markedly lowered. This suggests that the transition temperature $T_{c2}$ is independent of $T_{c1}$.   

\subsection{Resistivity in Pd film with Cu electrodes}
The Au electrode can form an impurity phase~\cite{27,28}, which may affect the superconducting transition of the Pd film. To examine this possibility, we measured the resistivity of the PdH$_{x}$ film with Cu electrodes deposited on the Pd film, as illustrated in Fig. A-1(b). The electrodes were not covered with the Pd film. 

The results are presented in Fig.A-4, where H absorption was performed at $T$ = 150 K in the same manner as for the Au electrodes. These qualitative features are consistent with the results obtained using Au electrodes. As the temperature decreases, the resistivity in the normal state decreases sharply at $T_{c1}$ = 2.30 K owing to the superconducting transition. Notably, although the residual resistivity below $T_{c1}$ is considerably smaller than that observed with the Au electrodes, it still remains after the transition, which is followed by the second transition at $T_{c2}$ = 0.71 K. The H concentration is estimated to be $x$ = 0.870. 
These results indicate that the emergence of the residual resistivity and the second transition are not due to the influence of the Au electrodes but are instead intrinsic features of superconducting PdH$_{x}$.


\begin{figure}[htbp]
\centering
\vspace*{-8cm} 
\hspace*{-0.5cm} 
\includegraphics [scale = 0.8, bb =0 0 450 620]{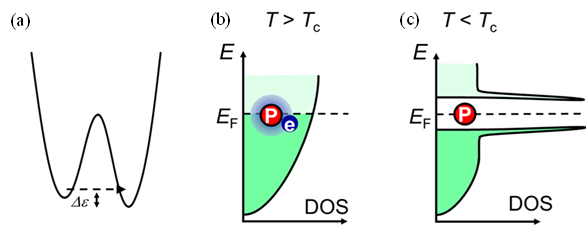}
\caption{Schematic illustration of hydrogen tunneling absorbed in a normal metal and a superconductor. (a) A schematic illustration of tunneling. Hydrogen can traverse an energy barrier between two interstitial sites, where $\Delta$$\varepsilon$ is the potential difference of the two wells. (b) In a normal metal, the electron density around the H atom increases due to the positive charge of the proton. Consequently, the tunneling proton migrates, dragging the electron cloud in diffusion. (c) In a superconductor, the emergence of an energy gap on the Fermi surface significantly reduces the interaction between the proton and conduction electrons, thereby enhancing the tunneling probability of proton during diffusion.}
\end{figure}

\begin{figure}
\centering
\vspace*{-3cm} 
\hspace*{2cm} 
\includegraphics [scale = 0.8, bb =0 0 660 510]{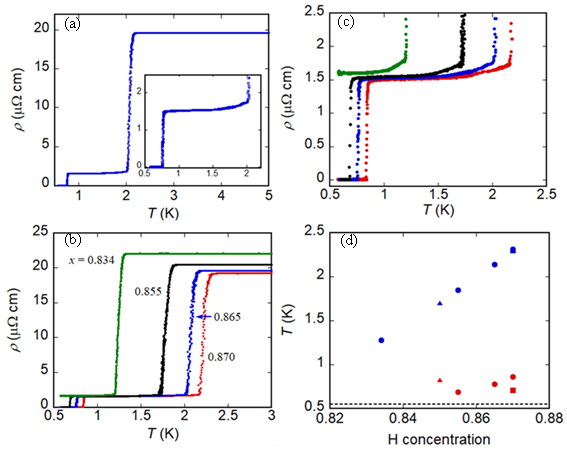}
\caption{Temperature dependence of the resistivity in PdHx films. (a) Temperature dependence of resistivity of the PdH$_{x}$ film, in which H atoms were absorbed at $T$ = 150 K. The superconducting transition occurs at $T_{c1}$ = 2.14 K, from which the H concentration is estimated to be $x$ = 0.865. The inset shows the low-temperature behavior. The resistivity drops steeply at $T_{c2}$ = 0.78 K, resulting in zero resistivity. (b) Temperature dependence of resistivity at $x$ = 0.870, 0.865, 0.855, and 0.834, which correspond to the curves at $T$ = 3 K from the bottom to top. The H concentration is varied by changing the absorption temperature, as described in Fig. A-2. (c) Enlarged view of the low-temperature region in (b). (d) H-concentration dependence of $T_{c1}$ and $T_{c2}$, marked by blue and red color, respectively. Circle, triangle, and square represent the transition temperatures in the Pd film depicted in (b), Pd film with a width of 10 ${\mu}$m in Fig. A-3 and Pd film with Cu electrodes in Fig. A-4, respectively. The dotted line indicates the minimum temperature in the measurements.  }
\end{figure}

 \begin{figure}
\centering
\vspace*{-3cm} 
\hspace*{2cm} 
\includegraphics [scale = 0.8, bb =0 0 600 540]{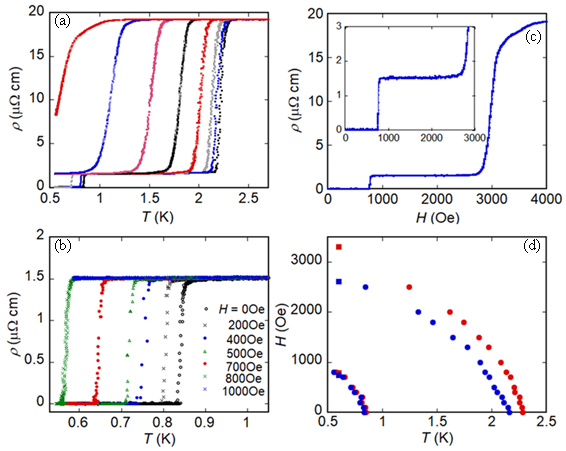}
\caption{ Magnetic field dependence of the resistivity in PdH$_{x}$ film. (a) Temperature dependence of the resistivity in PdH$_{0.870}$ film at $H$ = 0 Oe, 300 Oe, 500 Oe, 1000 Oe, 1500 Oe, 2000 Oe, 2500 Oe, and 3000 Oe. The transition curve at $T_{c1}$ is broadened with increasing the magnetic field. (b) Temperature dependence of the resistivity at $H$ = 0 Oe, 200 Oe, 400 Oe, 500 Oe, 700 Oe, 800, and 1000 Oe in the low temperature region. The steep drop of the resistivity is observed up to $H$ = 800 Oe. (c) Magnetic field dependence of the resistivity at $T$ = 0.6 K. The resistivity increases steeply at around $H$ = 750 Oe, followed by almost a constant resistivity up to $\sim$2500 Oe, as shown in the inset. Subsequently, it increases sharply at around 2700 Oe. (d) Temperature vs. magnetic field phase diagram of $T_{c1}$ and $T_{c2}$ for PdH$_{0.870}$. For both transitions, the onset (red circle) and endpoint (blue circle) are defined as 10 \% drop point from the normal (residual) resistivity and converging point to the baseline determined by the residual (zero) resistivity, respectively. Red and blue squares are derived from the magnetic dependence of the resistivity at $T$ = 0.6 K in (c), with transition points determined similarly.  }
\end{figure}

\begin{figure}
\centering
\vspace*{-0.5cm} 
\hspace*{4cm} 
\includegraphics [scale = 0.8, bb =0 0 660 545]{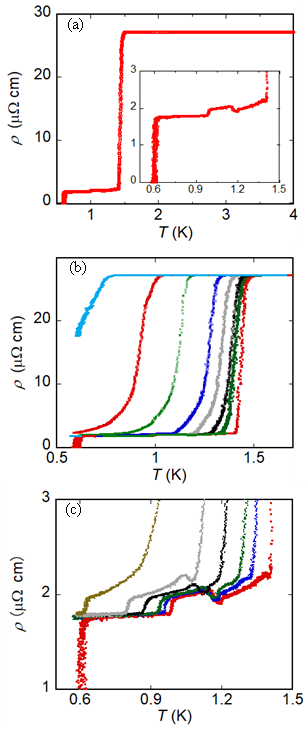}
\caption{Temperature dependence of the resistivity in PdD$_{x}$ film. (a) Temperature dependence of resistivity of the PdD$_{x}$ film, in which H atoms were absorbed at $T$ = 170 K. The superconducting transition occurs at $T_{c1}$ = 1.47 K, from which x is estimated to be 0.795. A residual resistivity remains after the superconducting transition and the second transition occurs at $T_{c2}$ $\sim$0.6 K. Additionally, a rectangular-shaped anomaly appears between $T$ = 0.95 K and 1.2 K. (b) Temperature dependence of the resistivity at $H$ = 0 Oe, 300 Oe, 500 Oe, 700 Oe, 1000 Oe, 1500 Oe, 2000 Oe, and 3000 Oe in PdD$_{0.795}$. The sample structure is identical to that used in Figs. 2 and 3. The transition curve at $T_{c1}$ is broadened with increasing the magnetic field, similar to the behavior observed in PdH$_{x}$. (c) Low-temperature resistivity as a function of temperature at $H$ = 0 Oe, 300 Oe, 500 Oe, 700 Oe, 1000 Oe, and 1500 Oe in the low temperature region. A rectangular-shaped structure between $T$ = 0.95 K and 1.2 K shifts to lower temperatures as the magnetic field increases.}
\end{figure}

\begin{figure}[h]
\renewcommand{\thefigure}{A-1}
\centering
\vspace*{-5cm} 
\hspace*{-4cm} 
\includegraphics [scale = 0.8, bb =0 0 300 500]{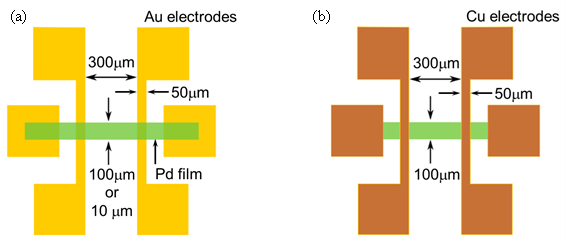}
\caption{Schematic illustration of Pd film. (a) Pd films with Au electrodes having a thickness of 20 nm and a width of 50 ${\mu}$m. The electrodes are prepared on a SiO$_{2}$/Si substrate with a 5-nm Ti layer deposited below the Au electrodes. A Pd film with a thickness of 100 nm and a width of 100 or 10  ${\mu}$m is deposited on the substrate with Au electrodes. The distance between the two voltage electrodes is 300  ${\mu}$m. (b) Pd film with Cu electrodes. After the Pd film with a thickness of 100 nm and width of 100  ${\mu}$m is deposited on a SiO$_{2}$/Si substrate by magnetron sputtering, the electrodes with a thickness of 100 nm and a width of 50  ${\mu}$m are deposited on the film by thermal evaporation. The distance between the two electrodes is 300  ${\mu}$m. }
\end{figure}

\begin{figure}[h]
\renewcommand{\thefigure}{A-2}
\centering
\hspace*{-1cm} 
\includegraphics [scale = 0.8, bb =0 0 300 500]{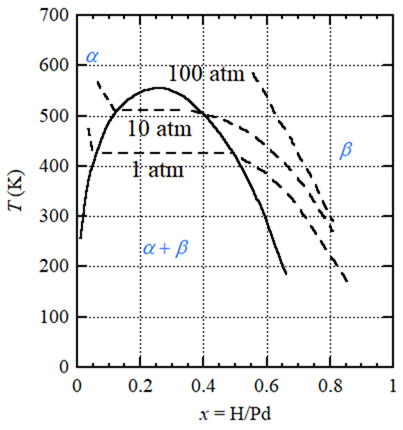}
\caption{The H concentration–pressure–temperature ($x–P–T$) phase diagram of Pd–H systems. The H concentration in Pd ($x$ = H/Pd) increases with decreasing temperature at the same H pressure.  }
\end{figure}

\begin{figure}[h]
\renewcommand{\thefigure}{A-3}
\centering
\hspace*{-2cm} 
\includegraphics [scale = 0.8, bb =0 0 300 500]{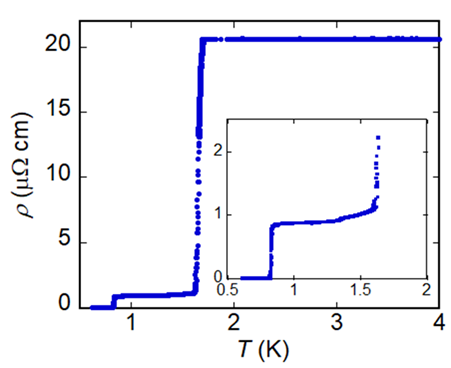}
\caption{Temperature dependence of resistivity in a Pd-hydride film for a small size. Temperature dependence of resistivity of a PdH$_{x}$ film with a width of 10 ${\mu}$m in which H atoms are absorbed at $T$ = 150 K. The superconducting transition and second transition occur at $T_{c1}$ = 1.70 K and at $T_{c2}$ = 0.83 K, respectively. The inset shows the low-temperature behavior.  }
\end{figure}

\begin{figure}[h]
\renewcommand{\thefigure}{A-4}
\centering
\hspace*{-2cm} 
\includegraphics [scale = 0.8, bb =0 0 300 500]{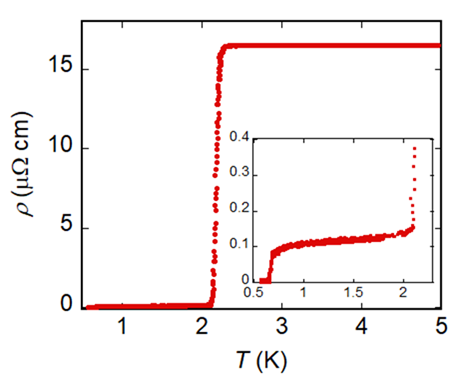}
\caption{Temperature dependence of resistivity in a Pd-hydride film with Cu electrodes. Temperature dependence of resistivity in a Pd film with Cu electrodes, where H atoms are loaded at $T$ = 150 K. The superconducting transition occurs at $T_{c1}$ = 2.30 K and finishes at 2.15 K. The inset shows the low-temperature behavior. As in the case of the Au electrodes, a residual resistivity remains, and a sharp drop occurs at $T_{c2}$ = 0.71 K.  }
\end{figure}

\end{document}